\documentstyle[11pt,moriond,epsfig]{article}

\def\Journal#1#2#3#4{{#1} {\bf #2}, #3 (#4)}

\def\PRL{\em Phys. Rev. Lett.}
\def\PRB{{\em Phys. Rev.} B}

\def\be{\begin{equation}}
\def\ee{\end{equation}}
\def\bea{\begin{eqnarray}}
\def\eea{\end{eqnarray}}

\def\gap{\scriptsize${\stackrel{\textstyle _>}{_\sim}} $\normalsize \ }
\def\lap{\scriptsize${\stackrel{\textstyle _<}{_\sim}} $\normalsize \ }
\begin{document}
%\vspace*{1cm}
\title{SUPPRESSION OF THE METALLIC BEHAVIOR IN TWO DIMENSIONS BY SPIN FLIP
SCATTERING}

\author{ DRAGANA POPOVI\'{C}, X. G. FENG }

\address{National High Magnetic Field Laboratory, Florida State
University, Tallahassee, FL 32310, USA}

\author{S. WASHBURN}

\address{Dept. of Physics and Astronomy, University of North Carolina at Chapel
Hill, Chapel Hill, NC 27599, USA}

\maketitle\abstracts{We study the effect of the disorder on the metallic
behavior of a two-dimensional electron system in silicon.  The temperature 
dependence of conductivity $\sigma (T)$ was measured for different values of 
substrate bias, which changes both potential scattering and the concentration 
of disorder-induced local magnetic moments.  We find that the latter has a 
much more profound effect on $d\sigma/dT$.  In fact, the data suggest that in 
the limit of $T\rightarrow 0$ the metallic behavior, as characterized by 
$d\sigma/dT < 0$, is suppressed by an arbitrarily small amount of spin flip 
scattering by local magnetic moments.}

\section{Introduction}

A metal-insulator transition (MIT) has been observed recently in a variety of
two-dimensional (2D) electron~\cite{Kravchenko,DP_MIT,AlAs_el,GaAs_el} and 
hole~\cite{SiGe_holes,GaAs_holes} systems but there is still no generally
accepted microscopic description of the 2D metallic phase.  Some of the 
relevant properties of the 2D metal include: (a) an increase of conductivity
$\sigma$ with decreasing temperature $T$ (i.~e. $d\sigma/dT < 0$) for carrier
densities $n_s > n_c$ ($n_c$ -- critical density); and (b) a suppression of the
$d\sigma/dT < 0$ behavior by magnetic field~\cite{Bfield}.  The latter suggests
the importance of the spin degrees of freedom, which can be probed further by
studying the effect of local magnetic moments on the transport properties of 
the conduction electrons.  In the experiment discussed below, the localized 
moments were induced by disorder and their number was varied in a controlled 
way.

The measurements were performed on a 2D electron system in Si 
metal-oxide-semiconductor field-effect transistors (MOSFETs).  In such a 
device, the disorder is due to the oxide charge scattering (scattering by 
ionized impurities randomly distributed in the oxide within a few \AA\, of the
interface) and to the roughness of the Si-SiO$_2$ interface~\cite{AFS}.  For a
fixed $n_s$, it is possible to change the disorder by applying the substrate
(back gate) bias $V_{sub}$.  In particular, the reverse (negative) $V_{sub}$
moves the electrons closer to the interface, which increases the disorder.  It
also increases the splitting between the subbands since the width of the 
triangular potential well at the interface is reduced by applying negative
$V_{sub}$.  Usually, only the lowest subband is occupied at low $T$, giving
rise to the 2D behavior.  In sufficiently disordered samples, however, the band
tails associated with the upper subbands may be populated even at low $n_s$
and act as additional scattering centers for mobile electrons in the lowest
subband.  Clearly, the 
negative $V_{sub}$ reduces this type of scattering by depopulating the upper
subband.  The effect of scattering by electrons localized deep in the tails of
the upper subband was first observed as an enhancement of the mobility $\mu$ at
low $n_s$ with negative $V_{sub}$~\cite{Alan}, and was subsequently studied in
more detail by other groups using different measurements and 
techniques~\cite{tails}.  Here we present a systematic study of $\sigma (T)$ 
as the disorder is varied using $V_{sub}$.  We show that scattering by 
electrons localized in the tail of the upper subband has a much more profound
effect on $d\sigma/dT$ than potential scattering due to oxide charges and
surface roughness.  This is attributed to spin flip scattering by electrons in
localized states that are singly populated due to a strong on-site Coulomb 
repulsion, and act as local magnetic moments.  For typical localization 
lengths of $\sim 100$~\AA\, in Si MOSFETs~\cite{AFS,Timp}, the on-site Coulomb
repulsion is $\sim 10$~meV.  Therefore, such states will be singly occupied at
low $n_s$~\cite{AFS}.

\section{Experimental Results}

Our measurements were carried out on n-channel Si MOSFETs with the oxide charge
density of $3\times 10^{10}$cm$^{-2}$.  Other details of the sample structure
have been given elsewhere~\cite{DP_MIT}.  For a fixed $V_{sub}$, $n_s$ was 
controlled 
by the gate voltage $V_g$ and determined in a standard fashion~\cite{AFS}.
$\sigma (V_g)$ was measured at temperatures $0.3 < T < 4.5$~K for $n_s$ of up
to $3\times 10^{12}$cm$^{-2}$ and for $-50\leq V_{sub}\leq {+1}$~V.  The effect
of $V_{sub}$ on $\mu$ at 4.2~K was found~\cite{icps_Feng} to be consistent with
earlier work~\cite{tails} and our interpretation.  In particular, for 
$n_s < n_{max}$ ($n_{max}\sim 5\times 10^{11}$cm$^{-2}$ is the density where 
$\mu$ reaches its maximum), an increase of $\mu$ is observed~\cite{icps_Feng} 
with the negative $V_{sub}$ as a result of the decreased scattering by local 
moments from the upper subband.  For $n_s > n_{max}$, $\mu$ decreases with
(negative) $V_{sub}$, consistent with the fact that surface roughness 
scattering is the dominant source of disorder in this range of 
$n_s$~\cite{AFS}.  For sufficiently high negative $V_{sub}$ ($-V_{sub} > 
35$~V), the 4.2~K mobility decreases with $V_{sub}$ for all $n_s$, suggesting
that the upper subband has been completely depopulated and that the further
increase in $V_{sub}$ leads only to increasing disorder due to potential
scattering from roughness at the Si-SiO$_2$ interface.

\begin{figure}[t]
%\rule{5cm}{0.2mm}\hfill\rule{5cm}{0.2mm}
%\vskip 2.5cm
%\rule{5cm}{0.2mm}\hfill\rule{5cm}{0.2mm}
\centerline{\psfig{figure=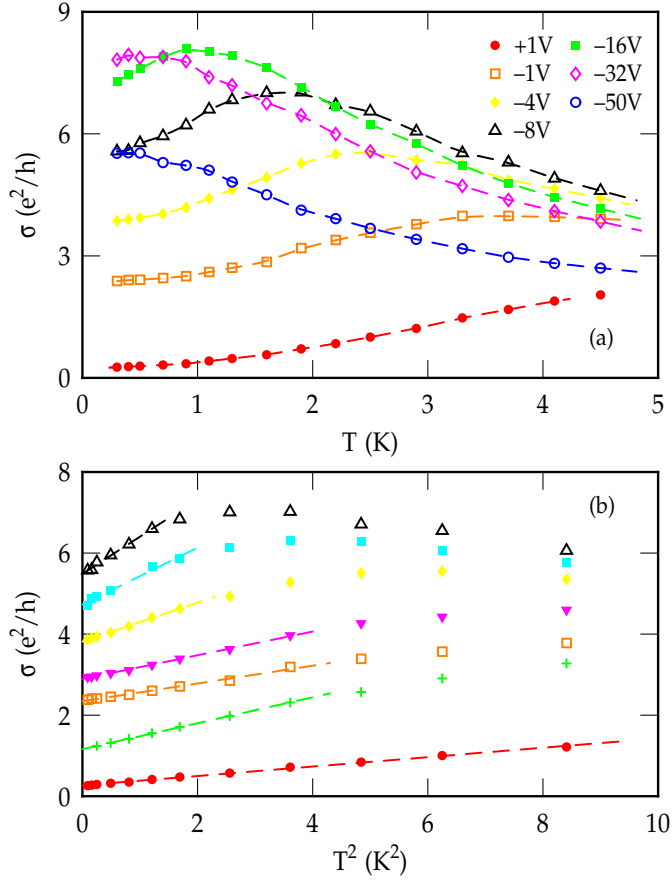,width=3.5in}}
\caption{Temperature dependence of the conductivity $\sigma$ for $n_s=2.0\times
10^{11}$cm$^{-2}$.  (a) The data are shown for different values of $V_{sub}$ as
given on the plot.  (b) The data are plotted vs. $T^2$, and shown for $V_{sub}=
+1, -0.5, -1, -2, -4, -6, -8$~V going from bottom to top.  Dashed lines are 
fits.
\label{fig1}}
\end{figure}

Fig.~\ref{fig1}(a) shows some typical results for $\sigma (T)$ at low $n_s$ as
a function of $V_{sub}$.  The metallic behavior, such that $d\sigma/dT < 0$, 
spreads out towards lower $T$ with the increasing negative $V_{sub}$, {\em 
i.~e.} as the scattering by local moments is reduced.  In other words, 
$\sigma (T)$
displays a maximum at $T=T_m$, such that $T_m$ shifts to lower $T$ with the
(negative) $V_{sub}$.  As ($-V_{sub}$) is increased beyond 35~V, {\em i.~e.} 
when the
upper subband is completely depopulated, the form of $\sigma (T)$ is no longer
very sensitive to changes in $V_{sub}$ even though the disorder due to 
potential scattering increases.  In addition, we note that $d\sigma/dT < 0$
behavior is more pronounced in the case where scattering by local moments is 
reduced even though the total disorder (4.2~K mobility) is larger (lower).
[See, for example, the data for $V_{sub}=-50$ and $-1$~V in 
Fig.~\ref{fig1}(a).]  This demonstrates clearly that scattering by
electrons localized in the tail of the upper subband has a much more profound,
and a {\em qualitatively different} effect on $d\sigma/dT$ than potential 
scattering due to oxide charges and surface roughness.  It is also 
qualitatively different from scattering among conduction electrons themselves,
which gives rise to a negative contribution to $d\sigma/dT$~\cite{Finkelstein}.
We find here that $d\sigma/dT > 0$ for $T<T_m$ and, in fact, $\sigma$ follows 
a $T^2$ form at the lowest $T$ [Fig.~\ref{fig1}(b)].  Such $\sigma (T)$ is 
often considered to be a signature of localized magnetic moments, and results 
from
the Kondo effect~\cite{Hewson}.  Here it represents a direct evidence for the 
existence of local moments in our samples.  A detailed study of this regime 
has been presented elsewhere~\cite{ourKondo}.  Fig.~\ref{fig1}(b) also shows 
that the range of $T$ ($T<T_m$) where local moments dominate transport becomes
smaller as their number is reduced by increasing negative $V_{sub}$.  

\begin{figure}
%\rule{5cm}{0.2mm}\hfill\rule{5cm}{0.2mm}
%\vskip 2.5cm
%\rule{5cm}{0.2mm}\hfill\rule{5cm}{0.2mm}
\centerline{\psfig{figure=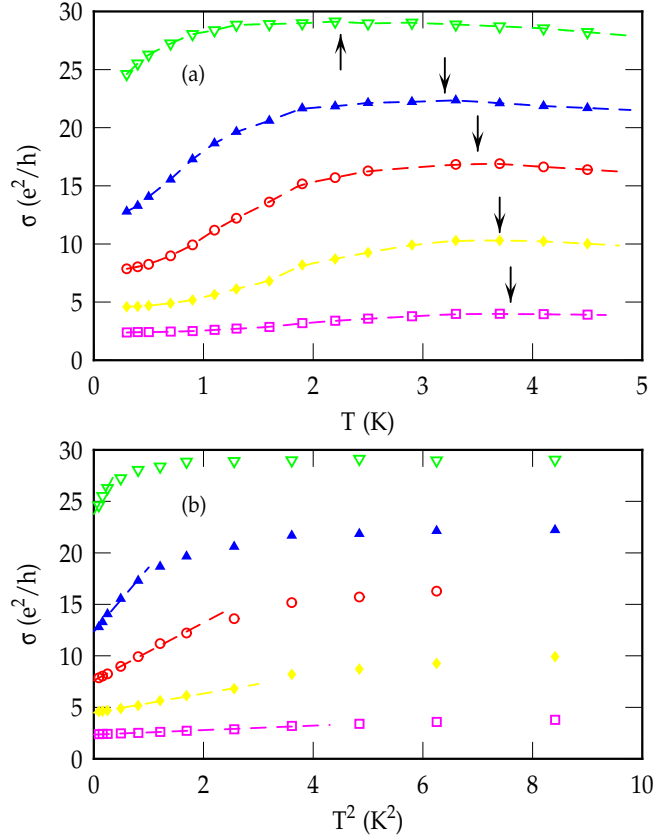,width=3.4in}}
\caption{$\sigma (T)$ for $V_{sub}=-1$~V, and $n_s=
2.0, 3.0, 4.0, 5.0, 7.0\times 10^{11}$cm$^{-2}$ going from bottom to top.
(a) The arrows indicate the peaks in $\sigma (T)$.  (b) The same data are 
plotted vs. $T^2$.  Dashed lines are fits.
\label{fig2}}
\end{figure}

For a fixed $V_{sub}$, $T_m$ shifts to lower $T$ with an increase in $n_s$
[Fig.~\ref{fig2}(a)], and the low $T$ regime where the $T^2$ behavior 
holds is correspondingly reduced [Fig.~\ref{fig2}(b)].  These data indicate 
that an increase in $n_s$ also reduces the number of local moments in the 
upper subband.  We note, however, that the change in the number of local
moments with $n_s$ becomes significant {\em only} when $n_s$\gap $n_{max}$, as 
discussed in more detail below.  

\begin{figure}
%\rule{5cm}{0.2mm}\hfill\rule{5cm}{0.2mm}
%\vskip 2.5cm
%\rule{5cm}{0.2mm}\hfill\rule{5cm}{0.2mm}
\centerline{\psfig{figure=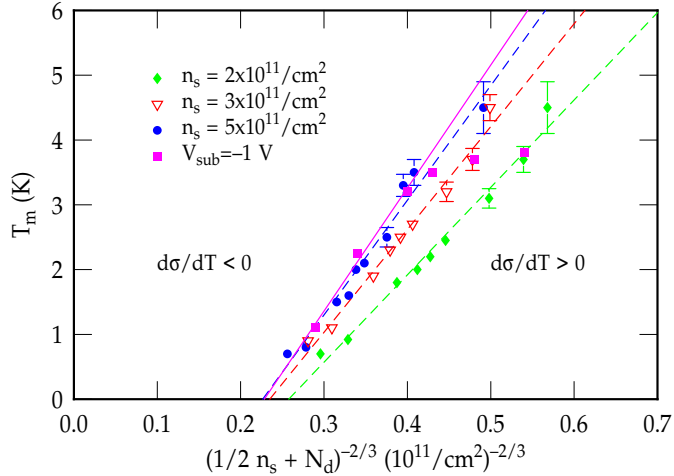,width=3.5in}}
\caption{Position of the maximum $T_m$ in $\sigma (T)$ for three different 
$n_s$ as a function of the inverse subband splitting (varied using $V_{sub}$).
The metallic behavior with $d\sigma/dT < 0$ is observable at $T>T_m$.  $T_m$ 
extrapolates to zero (dashed lines) for $V_{sub}\sim -40$~V, corresponding to 
the subband splitting of the order of 30~meV.  Squares show the data for 
$n_s=2.0,3.0,4.0,5.0,7.0,10.0\times 10^{11}$cm$^{-2}$ (right to left) and 
$V_{sub}=-1$~V.  The corresponding $T_m$ extrapolate to zero (solid line) at
the same value of the subband splitting.  (See also main text.)
\label{Tm}}
\end{figure}

The maximum position $T_m$ is shown in Fig.~\ref{Tm}.  The data are plotted
as a function of the inverse subband splitting for three values of $n_s$ 
in the metallic regime ($d\sigma/dT < 0$ for $T>T_m$).  The subband splitting
was controlled with $V_{sub}$ ($N_d$ is the depletion layer charge 
density, which increases with the reverse $V_{sub}$)~\cite{AFS}.  For each
density, $T_m$ extrapolates to zero for a finite value of the subband splitting
($V_{sub}\sim -40$~V).  This value is in agreement with the trend in the 4.2~K
mobility discussed above.  Fig.~\ref{Tm} also shows $T_m$ obtained for 
$V_{sub}=-1$~V by varying $n_s$ ({\em i.~e.} $V_g$) (the values are marked by 
arrows in Fig.~\ref{fig2}).  Increasing $V_g$ both raises $E_F$ and increases 
the subband splitting~\cite{AFS}.  Larger $E_F$ tends to increase the number 
of localized moments, but larger subband splitting decreases their number by
depopulating the upper subband.  It is also possible that the number of 
localized moments might be reduced at high $n_s$ because of improved screening
by the mobile electrons.  Fig.~\ref{Tm} shows that, for fixed $V_{sub}$ and 
$n_s$\lap $n_{max}$, $T_m$ only depends very weakly on $n_s$: it appears to 
decrease for some $V_{sub}$ ({\em e.~g.} $-1$~V), but to increase slightly for
other values of $V_{sub}$ ({\em e.~g.} $-2,-4,-8$~V).  This suggests that the 
two effects (raising $E_F$ and increasing the subband splitting) are 
comparable in size and that the number of localized moments for 
$n_s$\lap $n_{max}$ is, therefore, roughly constant.  On the other hand, there
is a rapid decrease of $T_m$ with $n_s$ for $n_s$\gap $n_{max}$.  Our data 
suggest (solid line in Fig.~\ref{Tm}) that this drop results from an increase
in the subband splitting and that other effects, such as screening, are not as
significant.  In fact, $T_m$ extrapolates to zero at
about the same value of the subband splitting as that obtained from 
$T_m(V_{sub})$ for a fixed $n_s$ (dashed lines in Fig.~\ref{Tm}): of the order
of 30 meV, which is consistent with earlier measurements of band 
tails~\cite{tails}.  Our results, therefore, show that the 2D metal with 
$d\sigma/dT < 0$ can exist at $T=0$ only in the absence of scattering by 
disorder-induced localized moments.  This is similar to the behavior observed 
in a magnetic field~\cite{Bfield}, and consistent with some theoretical 
models~\cite{Cast,Finkelstein,supercon}.

In the presence of scattering by localized moments, $d\sigma/dT < 0$ is 
observable at $T>T_m$.  For a fixed $V_{sub}$, $\sigma (n_s,T)$ for $T>T_m$ 
exhibits all of the properties of a 2D MIT~\cite{spinflip}.  We speculate that
localized moments might exist in other materials as well but that the 
corresponding $T_m$ might be too low to be experimentally accessible in 
high-mobility devices.

Back gate bias was used recently in a 2D hole system in 
GaAs/AlGaAs~\cite{spinsplit} to study the effect of the spin-splitting due to
the spin-orbit interaction and the inversion asymmetry of the confining
potential~\cite{Rashba}.  It was found that the magnitude of the $d\sigma/dT 
< 0$ behavior was reduced as the spin-splitting decreased, {\em i.~e.} as the 
confining potential became more symmetric.  In our samples, we observe the
opposite: the triangular confining potential becomes more symmetric with the
application of the reverse $V_{sub}$, and that is exactly when the $d\sigma/dT
< 0$ behavior appears.  Therefore, even if the effect of the spin-orbit
interaction exists in our samples, it does not drive the MIT.  Our conclusion
is consistent with the calculations~\cite{Zawadzki} that indicate that the
spin splitting in Si MOSFETs should be very close to zero, unlike that in some
other materials.

\section{Conclusion}

Our study shows that the 2D metal with $d\sigma/dT < 0$ can exist in the 
$T\rightarrow 0$ limit only in the absence of scattering by local magnetic
moments.  Our results emphasize the key role of the spin degrees of freedom in
the physics of the low density 2D electron system.

\section*{Acknowledgments}

The authors are grateful to V. Dobrosavljevi\'{c} and A. B. Fowler for helpful
discussions.  This work was supported by NSF Grant No. DMR-9796339.

\end{document}